\documentclass[final,5p,times,twocolumn]{elsarticle}

\usepackage{amssymb}
\usepackage{lipsum}
\usepackage{amsmath}

\usepackage{lineno}

\usepackage{tikz-feynhand}
\usepackage{hyperref}
\usepackage{physics}

\newcommand{\n}{\nonumber}
\newcommand{\st}{\sin\theta}
\newcommand{\ct}{\cos\theta}
\newcommand{\lam}{\lambda}
\newcommand{\sig}{\sigma}
\newcommand{\eps}{\epsilon}

\newcommand{\gam}{\gamma}
\newcommand{\bet}{\beta}

\journal{Physical Review D}

\begin{document}

\begin{frontmatter}



\title{W-boson pair production at lepton colliders in the Feynman-diagram gauge}


\author[1]{Hiroyuki Furusato}
\author[1,2,3]{Kentarou Mawatari}
\ead{mawatari@iwate-u.ac.jp}
\author[2]{Yutaro Suzuki}
\author[3]{Ya-Juan Zheng}

\address[1]{Graduate School of Science and Engineering, Iwate University, Morioka, Iwate 020-8550, Japan}
\address[2]{Graduate School of Arts and Sciences, Iwate University, Morioka, Iwate 020-8550, Japan}
\address[3]{Faculty of Education, Iwate University, Morioka, Iwate 020-8550, Japan}

\begin{abstract}
We calculate helicity amplitudes for $e^-e^+\to W^-W^+$ analytically in the Feynman-diagram (FD) gauge. 
We show that, unlike in the unitary gauge, 
there is no energy growth of the individual Feynman amplitudes for the longitudinally polarized W bosons, and 
the contributions from the associated Goldstone bosons are manifest even without taking the high-energy limit.
We also find that the physical distributions can be interpreted by the individual amplitudes in the FD gauge.  
\end{abstract}







\end{frontmatter}




\section{Introduction}\label{sec:intro}

It is known that gauge cancellation among amplitudes is an obstacle to numerical evaluations of cross sections as well as event generation, especially for the collinear phase space in QED and QCD, and for longitudinally polarized gauge-boson scattering in high energies in the electroweak theory. 

A recently proposed gauge fixing, Feynman-diagram (FD) gauge~\cite{Hagiwara:2020tbx,Chen:2022gxv,Chen:2022xlg}, is promising for high-energy event simulations not only for the standard model (SM) processes but also for those beyond the SM~\cite{Hagiwara:2024xdh}.
So far, all the results shown in refs.~\cite{Hagiwara:2020tbx,Chen:2022gxv,Chen:2022xlg,Hagiwara:2024xdh} are numerical and indicate that subtle gauge cancellation among the interfering amplitudes can be avoided. 

In this paper, we revisit the $e^-e^+\to W^-W^+$ process in the SM in order to study the analytic structure of the helicity amplitudes in the FD gauge for the first time.

The process has been thoroughly studied both theoretically, e.g.\ in refs.~\cite{Hagiwara:1986vm,Gounaris:1996rz,Beenakker:1996kt}, and experimentally~\cite{ALEPH:2013dgf} in the LEP era. 
The process for longitudinally polarized W bosons is also often discussed in quantum-field-theory textbooks to demonstrate gauge cancellation and the Goldstone boson equivalence theorem; see e.g.\ Chapter~21 in Peskin and Schroeder~\cite{Peskin:1995ev}. 
Moreover, W-boson pair production is very important for the precision test of the electroweak theory in future high-energy lepton colliders,
such as the ILC~\cite{ILC:2013jhg,Fujii:2017vwa,ILCInternationalDevelopmentTeam:2022izu},
CEPC~\cite{CEPCStudyGroup:2018ghi}, and FCC-ee~\cite{FCC:2018evy}.

We note that most of the tree-level calculations and discussions have been conventionally done in the unitary (U) gauge. 
In ref.~\cite{Chen:2022gxv}, the process $e^-e^+\to W^-W^+\to \ell^-\bar\nu_\ell\ell'^+\nu_{\ell'}$ is studied numerically 
to demonstrate the calculation in the FD gauge by comparing with that in the U gauge.
Here, we calculate the $e^-e^+\to W^-W^+$ scattering amplitudes analytically in the FD gauge, and 
show that gauge cancellation among the amplitudes for longitudinally polarized W bosons in the U gauge at high energies is absent in the FD gauge
since there is no energy growth of the individual FD-gauge amplitudes.

After introducing the FD gauge in Sect.~\ref{sec:fd}, 
we present the helicity amplitudes for $e^-e^+\to W^-W^+$ both in the FD and U gauges in Sect.~\ref{sec:amp}.
In Sect.~\ref{sec:xsec} we show the numerical results for the total and differential cross sections 
to discuss the contributions from the individual Feynman amplitudes in the two gauges. 
Section~\ref{sec:summary} is devoted to a summary.

\section{Feynman-diagram gauge}\label{sec:fd}

In this section, we briefly introduce ingredients necessary to calculate helicity amplitudes for $e^-e^+\to W^-W^+$ in the FD gauge.

In the FD gauge, the propagator of the gauge bosons is obtained from the light-cone gauge~\cite{Chen:2022xlg}, 
where the gauge vector is chosen along the opposite direction of the gauge-boson three momentum
\begin{align}
n^\mu = ( {\rm sgn}(q^0), -\vec{q}/|\vec{q}| )\ .
\label{eq:FD}
\end{align}
Note that we take $n^\mu = (1,0,0,-1)$ when $|\vec{q}|=0$ and $q^0>0$.
The propagator of a massless gauge boson is given by~\cite{Hagiwara:2020tbx}
\begin{align}
G_{\mu\nu}(q) = \frac{i}{q^2+i\epsilon} \qty( -g_{\mu\nu} 
+ \frac{q_{\mu}n_{\nu}+n_{\mu}q_{\nu}  }{n\cdot q } )\ , 
\label{fdpropagator1}
\end{align}
while that of a massive gauge boson is combined with the associated Goldstone boson
and formed as a five-dimensional propagator with its mass $m$~\cite{Chen:2022gxv},
\begin{align}
G_{MN}(q) = \frac{i}{q^2-m^2+i\epsilon} 
  \begin{pmatrix}
	  -g_{\mu\nu} 
	  + \dfrac{q_{\mu}n_{\nu}+n_{\mu}q_{\nu}  }{n\cdot q } & i \dfrac{m\, n_\mu}{n\cdot q} \\
	  -i \dfrac{m\, n_\nu}{n\cdot q} & 1 \\
  \end{pmatrix}	
\ , 
\label{fdpropagator2}
\end{align}
with $M,N=0$ to 4, where 0 to 3 are the Lorentz indices $\mu,\nu$.

The polarization vector (or wave function) for massive gauge bosons with the helicity $\lam\,(=\pm1,0)$ in the FD gauge includes the associated Goldstone boson, and is given as a five-component vector by~\cite{Chen:2022gxv}
\begin{align}
 \eps^M(q,\pm)&=(\eps^\mu(q,\pm),\,0)\ , 
 \label{fdpolT}\\
 \eps^M(q,0)&=(\tilde\eps^\mu(q,0),\,i)\ ,
 \label{fdpolL}
\end{align}
with the reduced polarization vector~\cite{Hagiwara:2020tbx,Chen:2022gxv,Chen:2022xlg}
\begin{align}
 \tilde\eps^\mu(q,0)&=\eps^\mu(q,0)-\frac{q^\mu}{Q}
  =-{\rm sgn}(q^2)\frac{Q\,n^\mu}{n\cdot q} \ ,
 \label{e0tilde}
\end{align}
where $\eps^\mu(q,\lam)$ is the ordinary polarization vector and $Q=\sqrt{|q^2|}$.

It is well known that the ordinary polarization vector with $\lam=0$, i.e.\  
the longitudinally polarized gauge boson, behaves badly at high energies, as shown below.
For a massive gauge boson with its momentum
\begin{align}
 q^\mu=(E, 0 , 0, q)\ ,
 \label{mom}
\end{align}
the transverse polarization vector is given by 
\begin{align}
 \eps^\mu(q,\pm)=\frac{1}{\sqrt{2}}(0,\mp1,-i,0) \ ,
\end{align}
while the longitudinal polarization vector is
\begin{align}
 \eps^\mu(q,0)=\frac{1}{m}(q,0,0,E)
 =\gam\,(\bet,0,0,1) \ ,
 \label{e0_u}
\end{align}
where 
\begin{align}
 \bet=\frac{q}{E}=\sqrt{1-\frac{m^2}{E^2}}\ ,\quad \gam=\frac{E}{m}=\frac{1}{\sqrt{1-\bet^2}}\ .
 \label{boost}
\end{align}
This $\gam$ factor gives rise to the energy growth of scattering amplitudes at high energies ($\bet\to1$, $\gam\to\infty$), which might cause violation of unitarity. 

In contrast, the reduced polarization vector for $\lam=0$
 in eq.~\eqref{e0tilde} 
 with the momentum \eqref{mom} is given as
\begin{align}
 \tilde\eps^\mu(q,0)=\frac{1}{\gam(1+\bet)}\,(-1,0,0,1) \ ,
 \label{e0_fd}
\end{align}
which vanishes in the high-energy limit.

\section{Helicity amplitudes}\label{sec:amp}

In this section we present helicity amplitudes in the FD gauge for the process
\begin{align}
 e^-(k,\sigma)+e^+({\bar{k}},\bar{\sigma})\rightarrow W^-(p,\lambda)+W^+(\bar{p},\bar{\lambda})
\end{align}
in the SM.
The four-momenta ($k$, $\bar k$, $p$, $\bar p$) and the helicities ($\sig$, $\bar\sig$, $\lam$, $\bar\lam$) are defined in the $e^-e^+$ collision frame
\begin{align}
 k^\mu&=\frac{\sqrt{s}}{2}(1,0,0,1)\ , \n\\
 \bar{k}^\mu&=\frac{\sqrt{s}}{2}(1,0,0,-1)\ , \n\\
 p^\mu&=\frac{\sqrt{s}}{2}(1,\beta\sin\theta,0,\beta\cos\theta)\ , \n\\
 \bar{p}^\mu&=\frac{\sqrt{s}}{2}(1,-\beta\sin\theta,0,-\beta\cos\theta)\ ,
 \label{kin}
\end{align}
with the center-of-mass energy $\sqrt{s}$ and the scattering angle $\theta$ for $W^-$ with respect to the $e^-$ momentum direction. 
The boost factors in eq.~\eqref{boost} in this frame are
\begin{align}
 \bet=\sqrt{1-\frac{4m_W^2}{s}}\ ,\quad \gam=\frac{\sqrt{s}}{2m_W}\ .
\end{align}

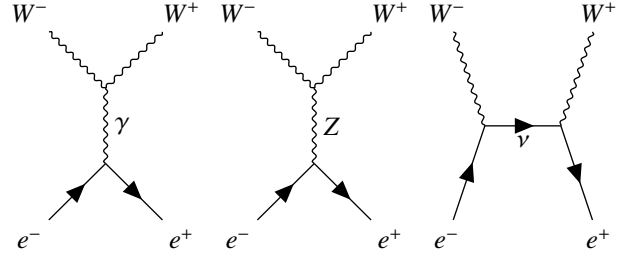
\begin{figure}
\centering
\begin{tabular}{c}
 
\begin{minipage}[b]{0.3\linewidth}
\centering
\begin{tikzpicture}
\begin{feynhand}
\vertex (a) at (0,0); \vertex (b) at (1,1) {$W^+$}; \vertex (c) at (-1,1) {$W^-$};
\vertex (d) at (0,-1); \vertex (e) at (-1,-2) {$e^-$}; \vertex (f) at (1,-2) {$e^+$};
\propag [photon] (a) to (b); \propag [photon] (a) to (c); \propag [photon] (a) to[edge label=$\gamma$] (d);
\propag [fermion] (e) to (d); \propag[anti fermion] (f) to (d);
\end{feynhand}
\end{tikzpicture}
\end{minipage} 

\begin{minipage}[b]{0.3\linewidth}
\centering
\begin{tikzpicture}
\begin{feynhand}
\vertex (a) at (0,0); \vertex (b) at (1,1) {$W^+$}; \vertex (c) at (-1,1) {$W^-$};
\vertex (d) at (0,-1); \vertex (e) at (-1,-2) {$e^-$}; \vertex (f) at (1,-2) {$e^+$};
\propag [photon] (a) to (b); \propag [photon] (a) to (c); \propag [photon] (a) to[edge label=$Z$] (d);
\propag [fermion] (e) to (d); \propag[anti fermion] (f) to (d);
\end{feynhand}
\end{tikzpicture}
\end{minipage} 

\begin{minipage}[b]{0.3\linewidth}
\centering
\begin{tikzpicture}
\begin{feynhand}
\vertex (a) at (-0.5,-0.5); \vertex (b) at (1,1) {$W^+$}; \vertex (c) at (-1,1) {$W^-$};
\vertex (d) at (0.5,-0.5); \vertex (e) at (-1,-2) {$e^-$}; \vertex (f) at (1,-2) {$e^+$};
\propag [photon] (d) to (b); \propag [photon] (a) to (c); \propag [fermion] (a) to[edge label'=$\nu$] (d);
\propag [fermion] (e) to (a); \propag[anti fermion] (f) to (d);
\end{feynhand}
\end{tikzpicture}
\end{minipage} 

\end{tabular}
\caption{Feynman diagrams for $e^-e^+\to W^-W^+$.}
\label{fig:diagram}
\end{figure}

There are three Feynman diagrams both in the FD and U gauges: $s$-channel photon ($\gam$) and Z-boson exchange, 
and $t$-channel neutrino exchange, 
\begin{align}
 {\cal M} =\sum_i {\cal M}_i = {\cal M}_\gam + {\cal M}_Z + {\cal M}_\nu \ ,
\label{totalamp} 
\end{align}
depicted in Fig.~\ref{fig:diagram}.
We note that, although the Feynman diagrams look identical both in the FD and U gauges, 
the weak-boson lines in the FD gauge implicitly include the associated Goldstone bosons
forming the $5\times5$ component propagator in eq.~\eqref{fdpropagator2} and the five-component polarization vectors in eqs.~\eqref{fdpolT} and \eqref{fdpolL},
as introduced in the previous section.
In the end of this section, we see that the Goldstone-boson contribution is manifest in FD-gauge amplitudes.

\begin{table}[b]
  \centering
\caption{Coupling and propagator factors for each amplitude.
 $s_W\equiv\sin\theta_W$ is the weak mixing angle in the SM.}
\label{tab:coup}  
\begin{tabular}{c|ccc}
  \hline
  $i$ & $\gam$ & $Z$ & $\nu$ \\ \hline
  $c_i$ & 1 & $s^{-2}_W\left(-\frac{1}{2}\delta_{\sigma,-1}+s^2_W\right)$
  & $s^{-2}_W\delta_{\sigma,-1}$ \\  
  $P_i(\theta)^{-1}$ & 1 & $1-m_Z^2/s$ & $1+\beta^2-2\beta\ct$\\
  \hline 
\end{tabular}
\end{table}

\begin{table*}
\centering
{\renewcommand\arraystretch{1.68}
  \caption{Reduced helicity amplitudes ${\tilde{\cal M}}_i{}_{}^{\lam\bar\lam}(\theta)$ in eq.~\eqref{amp} in the FD gauge.}
  \label{tab:amp_fd}
  \begin{tabular}{rcccc}  \hline
    $\Delta\lambda$ &    $(\lambda,\bar{\lambda})$ & ${\tilde{\cal M}}_\gam{}_{}^{\lam\bar\lam}(\theta)$ & ${\tilde{\cal M}}_Z{}_{}^{\lam\bar\lam}(\theta)$ & ${\tilde{\cal M}}_\nu{}_{}^{\lam\bar\lam}(\theta)$ \\ 
\hline \hline
    $0$&$(0,0)$ &        $\dfrac{1}{\gam^2}\dfrac{3+\beta}{(1+\bet)^2}+1$ &              
	$-\dfrac{1}{\gam^2}\dfrac{3+\beta}{(1+\bet)^2}-\dfrac{s^2_W}{c^2_W}\qty(\dfrac{\beta}{2s^2_W}-1)$    & 
	$-\dfrac{1}{\gam^2}\dfrac{2}{(1+\bet)^2}(1+\ct)$ \\ \hline
    $+1$&$(+,0),(0,-)$ &  $\dfrac{1}{2\gam}\qty(\dfrac{3-\beta}{1+\bet}+1)$  & 
	$-\dfrac{1}{2\gam}\qty(\dfrac{3-\beta}{1+\bet}-\dfrac{s^2_W}{c^2_W})$  & 
	$-\dfrac{1}{\gam}\dfrac{2}{1+\bet}(1+\ct)$ \\ \hline
    $-1$&$(0,+),(-,0)$ & $\dfrac{1}{2\gam}\qty(\dfrac{3-\beta}{1+\bet}+1)$ & $-\dfrac{1}{2\gam}\qty(\dfrac{3-\beta}{1+\bet}-\dfrac{s^2_W}{c^2_W})$ & $\dfrac{1}{\gam}\dfrac{2}{1+\bet}(\beta-\ct)$ \\ \hline
    $0$&$(\pm,\pm)$ &   
    $-\beta$ &                            $\beta$ & $\beta-\ct$ \\ \hline
    $\pm2$&$(\pm,\mp)$ &        0 &    0 & $-\sqrt{2}$ \\ \hline
  \end{tabular}
  }
\end{table*}

\begin{table*}
\centering
{\renewcommand\arraystretch{1.68}
  \caption{Same as Table~\ref{tab:amp_fd}, but in the U gauge.}
  \label{tab:amp_u}
  \begin{tabular}{rcccc}  \hline
    $\Delta\lambda$ &    $(\lambda,\bar{\lambda})$ & ${\tilde{\cal M}}_\gam{}_{}^{\lam\bar\lam}(\theta)$ & ${\tilde{\cal M}}_Z{}_{}^{\lam\bar\lam}(\theta)$ & ${\tilde{\cal M}}_\nu{}_{}^{\lam\bar\lam}(\theta)$ \\ \hline \hline
    $0$&$(0,0)$ &        $-2\gamma^2\beta+\beta$ &  $2\gamma^2\beta-\beta$   & $2\gamma^2(\bet-\ct)-\bet$ \\ \hline
    $+1$&$(+,0),(0,-)$ &  $-2\gam\bet$ & $2\gam\bet$  &  $2\gamma(\beta-\ct)-\dfrac{1}{\gam}$   \\ \hline
    $-1$&$(0,+),(-,0)$ & $-2\gam\bet$ & $2\gam\bet$  &  $2\gamma(\beta-\ct)+\dfrac{1}{\gam}$   \\ \hline
    $0$&$(\pm,\pm)$ &  
    $-\beta$ &                            $\beta$ & $\beta-\ct$ \\ \hline
    $\pm2$&$(\pm,\mp)$ &        0 &    0 & $-\sqrt{2}$ \\ \hline
  \end{tabular}
  }
\end{table*}

Each helicity amplitude ($i=\gam,Z,\nu$) is written as~\cite{Hagiwara:1986vm}
\begin{align}
 {{\cal M}_i}_{\sig}^{\lam\bar\lam}
 =\sqrt{2}\,e^2\,c_i\, {\tilde{\cal M}}_i{}_{}^{\lam\bar\lam}(\theta)\,
  \varepsilon\, d^{J_0}_{\Delta\sigma,\Delta\lambda}(\theta)\,P_i(\theta)\ ,
  \label{amp}
\end{align}
where the coupling factor $c_i$ and the propagator factor $P_i(\theta)$ are given in Table~\ref{tab:coup},
$\varepsilon=\Delta\sigma(-1)^{\bar{\lambda}}$ is a sign factor, 
$\Delta\sigma=(\sigma-\bar{\sigma})/2$, $\Delta\lambda=\lambda-\bar{\lambda}$, $J_0=\textrm{max}(|\Delta\sigma|,|\Delta\lambda|)$,
and $d^{J_0}_{\Delta\sigma,\Delta\lambda}(\theta)$ is the $d$ function; see, e.g.\ ref.~\cite{ParticleDataGroup:2022pth} for the explicit forms.
We note that, since we neglect the electron mass, $\bar\sig=-\sig$, i.e.\ $\Delta\sig=+1$ or $-1$ and hence $J_0=1$ or 2.
All the above factors in eq.~\eqref{amp} except the reduced helicity amplitudes ${\tilde{\cal M}}_i{}_{}^{\lam\bar\lam}(\theta)$ are common both in the FD and U gauges.
We present ${\tilde{\cal M}}_i{}_{}^{\lam\bar\lam}(\theta)$ in the FD gauge in Table~\ref{tab:amp_fd},
while those in the U gauge are shown in Table~\ref{tab:amp_u} for comparison; see also Table~3 in ref.~\cite{Hagiwara:1986vm}.

Although it is not so obvious at a glance for $\lam=0$ and/or $\bar\lam=0$, the sum of the three amplitudes~\eqref{totalamp} in the FD gauge agrees with that in the U gauge for each helicity combination
because of the gauge invariance of the helicity amplitudes.

\vspace*{-0.5mm}
On the other hand, each amplitude~\eqref{amp} for the longitudinally polarized W bosons, $\lam=0$ and/or $\bar\lam=0$, is rather different between the two gauges.
For the $\lam=\bar\lam=0$ case, the first term of each amplitude in the FD gauge is proportional to $\gam^{-2}$, 
while that in the U gauge is proportional to $\gamma^2$.
Similarly, for the case that one of the W bosons is longitudinally polarized, i.e.\ $\Delta\lam=\pm1$,  
each amplitude in the FD gauge is proportional to $\gam^{-1}$, 
while the leading term of each amplitude in the U gauge is proportional to $\gamma$.
For the case that both W bosons are transversely polarized, 
each amplitude is identical in the two gauges,
 and does not depend on $\gam$.
These $\gam$ dependences are dictated by the longitudinal polarization vectors, 
eq.~\eqref{e0_fd} in the FD gauge and eq.~\eqref{e0_u} in the U gauge,
which behave completely opposite.
Therefore, in the high-energy limit $\beta\to1$, or $\gam\to\infty$, the amplitudes for longitudinally polarized W bosons behave
 rather differently 
in the two gauges,
which will be shown numerically in the next section.

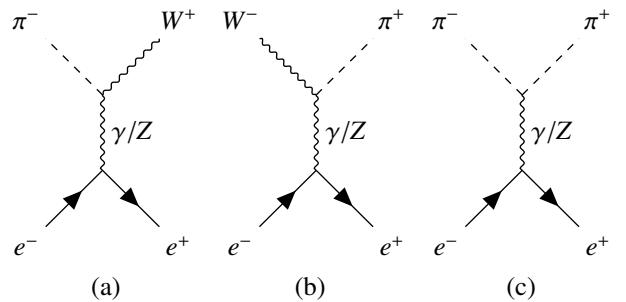
\begin{figure}
\centering
\begin{tabular}{c}
 
\begin{minipage}[b]{0.3\linewidth}
\centering
\begin{tikzpicture}
\begin{feynhand}
\vertex (a) at (0,0); \vertex (b) at (1,1) {$W^+$}; \vertex (c) at (-1,1) {$\pi^-$};
\vertex (d) at (0,-1); \vertex (e) at (-1,-2) {$e^-$}; \vertex (f) at (1,-2) {$e^+$};
\propag [photon] (a) to (b); \propag [scalar] (a) to (c); \propag [photon] (a) to[edge label=$\gamma/Z$] (d);
\propag [fermion] (e) to (d); \propag[anti fermion] (f) to (d);
\end{feynhand}
\end{tikzpicture}
\end{minipage} 

\begin{minipage}[b]{0.3\linewidth}
\centering
\begin{tikzpicture}
\begin{feynhand}
\vertex (a) at (0,0); \vertex (b) at (1,1) {$\pi^+$}; \vertex (c) at (-1,1) {$W^-$};
\vertex (d) at (0,-1); \vertex (e) at (-1,-2) {$e^-$}; \vertex (f) at (1,-2) {$e^+$};
\propag [scalar] (a) to (b); \propag [photon] (a) to (c); \propag [photon] (a) to[edge label=$\gamma/Z$] (d);
\propag [fermion] (e) to (d); \propag[anti fermion] (f) to (d);
\end{feynhand}
\end{tikzpicture}
\end{minipage} 

\begin{minipage}[b]{0.3\linewidth}
\centering
\begin{tikzpicture}
\begin{feynhand}
\vertex (a) at (0,0); \vertex (b) at (1,1) {$\pi^+$}; \vertex (c) at (-1,1) {$\pi^-$};
\vertex (d) at (0,-1); \vertex (e) at (-1,-2) {$e^-$}; \vertex (f) at (1,-2) {$e^+$};
\propag [scalar] (a) to (b); \propag [scalar] (a) to (c); \propag [photon] (a) to[edge label=$\gamma/Z$] (d);
\propag [fermion] (e) to (d); \propag[anti fermion] (f) to (d);
\end{feynhand}
\end{tikzpicture}
\end{minipage} 

\end{tabular}
(a)\hspace*{2.2cm} (b) \hspace*{2.2cm} (c)
\caption{Goldstone-boson ($\pi^\mp$) contributions to $e^-e^+\to W^-W^+$.}
\label{fig:diagram_gb}
\end{figure}

\vspace*{-0.5mm}
Before turning to numerical results, we remark on the Goldstone-boson contributions to the amplitudes in the FD gauge. 
In contrast to the U gauge, where the Goldstone bosons do not present explicitly, 
the contribution of Goldstone bosons is manifest in the FD gauge even without taking the high-energy limit.  
The second term of the $s$-channel $\gam/Z$ exchange amplitudes for $\lam=0$ and/or $\bar\lam=0$ in Table~\ref{tab:amp_fd} is exactly a footprint of the Goldstone bosons.
In Fig.~\ref{fig:diagram_gb}, we explicitly show the Goldstone-boson diagrams extracted from the diagrams in the FD gauge in Fig.~\ref{fig:diagram}.
We refer to Table~2 in ref.~\cite{Chen:2022gxv} for the couplings of $\pi^\mp$-$W^\pm$-$\gam/Z$ and $\pi^-$-$\pi^+$-$\gam/Z$.  
For $\lam=\bar\lam=0$, all three amplitudes in Fig.~\ref{fig:diagram_gb} contribute,
because both polarization vectors have the fifth component in eq.~\eqref{fdpolL}, 
representing the Goldstone boson component of the five-component weak boson.
Only diagram (a) gives nonvanishing contribution for $\lam=0$ and $\bar\lam=\pm1$,
while only diagram (b) contributes for $\lam=\pm1$ and $\bar\lam=0$.
For the $\lam=\bar\lam=0$ case 
in Table~\ref{tab:amp_fd}, as energy increases, the first term 
of the $\gam/Z$ exchange amplitudes
rapidly falls as $\gam^{-2}$ and the second term becomes dominant. 
This behavior is literally consistent with the Goldstone boson equivalence theorem.  
We note that the $Z$-associated $\pi^0$ contribution is absent since we neglect the electron mass.
Similarly, the $\pi^\pm$ do not couple to the massless fermion line in the $\nu$-exchange diagram.

\begin{figure*}
\centering
\includegraphics[height=0.272\textheight]{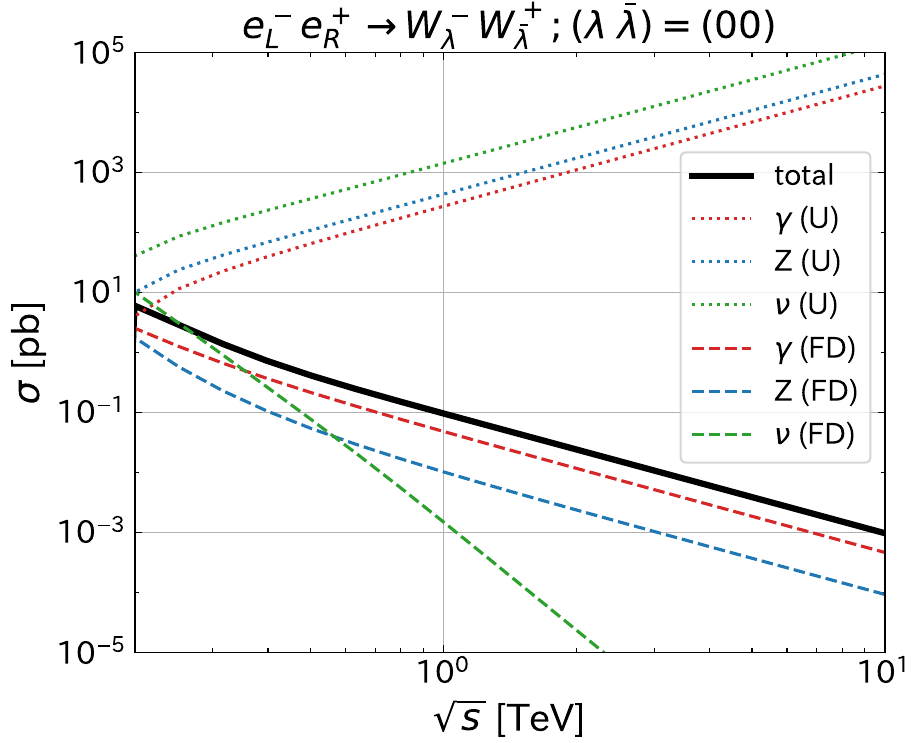}\qquad
\includegraphics[height=0.272\textheight]{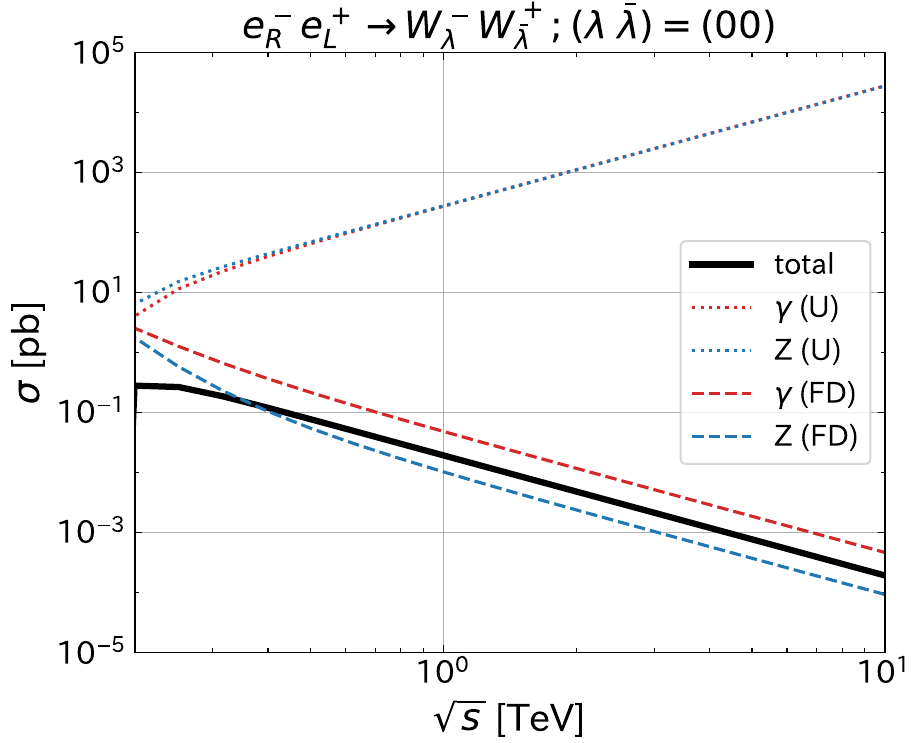}
\caption{Total cross section of $e^-e^+\rightarrow W^-W^+$ for $(\lambda,\bar{\lambda})=(0,0)$ as a function of the collision energy
with the left-handed electron (left) and the right-handed electron (right).
The solid line denotes the total cross section, while dashed (dotted) lines show contributions from the absolute value square of the individual amplitude in the FD (U) gauge.}
\label{fig:xsec_00}
\end{figure*}

\section{Cross sections}\label{sec:xsec}

In this section we present the total and differential cross sections for $e^-e^+\to W^-W^+$.
Only the sum of all the relevant amplitudes is gauge invariant, and its absolute value square gives a physical cross section.
Nevertheless, we also show contributions from the absolute value square of the individual Feynman amplitude
in order to compare the behaviors of each amplitude in the FD and U gauges.
In the following we focus on the case of longitudinally polarized W bosons in the final state ($\lam=0$ and/or $\bar\lam=0$) 
to discuss the difference between the two gauges. 

Figure~\ref{fig:xsec_00} shows the total cross section of $e^-e^+\rightarrow W^-W^+$ for $(\lambda,\bar{\lambda})=(0,0)$ as a function of the collision energy from 200~GeV up to 10~TeV, 
where we simply assume 100\% polarized beams as the left-handed (right-handed) electron in the left (right) panel.
The black solid line denotes the physical total cross section. 
Red, blue, and green dashed lines show contributions from the square of each $\gam$, $Z$, and $\nu$ amplitude, respectively, in the FD gauge.
Similarly, the dotted lines denote the U-gauge amplitudes. 

The total cross section is identical between the two gauges, showing gauge invariance of the sum of all the amplitudes,
and falls as $1/s$ for $s\gg m_Z^2$. 

On the other hand, all three individual amplitude squares in the U gauge grow with energy, dictated by the $\gam^2$ factor of the amplitudes in Table~\ref{tab:amp_u}.
There is a slight constant difference between the photon (red dotted) and $Z$ (blue dotted) contributions when $s\gg m_Z^2$ in the left panel,
while the two lines are overlapped in the right panel. 
Since the reduced $\gam$ and $Z$ amplitudes in Table~\ref{tab:amp_u} are identical except for the relative sign,
these can be explained by the gauge couplings of electrons and the propagator factors in Table~\ref{tab:coup}. 
This indicates that 
artificial gauge cancellation among the individual contributions with the order of $\big(s/m_Z^2\big)$ is required at the amplitude level to get the physical cross section, 
e.g.\ about $O(10^8)$ cancellation at $\sqrt{s}=10$~TeV as seen in Fig.~\ref{fig:xsec_00}.
This is exactly a problem of numerical evaluation for scattering processes at future high-energy colliders which we have encountered,
since matrix-element event generators normally evaluate each Feynman amplitude for a fixed helicity combination, sum up all of them, and square it.

In the FD gauge, in contrast, 
the photon and $Z$ contributions fall as $1/s$ in high energies, the same as the total cross section, while the $\nu$ contribution falls as $1/s^3$. 
There is absolutely no artificial cancellation among the relevant amplitudes. 
We clearly see that the Goldstone-boson contributions, i.e.\ the second term of the $\gam/Z$ amplitudes in Table~\ref{tab:amp_fd}, become dominant when $\gam^{-2}=4m_W^2/s\ll1$.
Similar to the relation between the photon and $Z$ amplitudes in the U gauge, the first terms of the photon and $Z$ amplitudes in the FD gauge in Table~\ref{tab:amp_fd}, i.e.\ the pure gauge-boson terms, are identical except for the relative sign.
For the Goldstone-boson terms, on the other hand, the $Z$ amplitudes are suppressed by $s_W^2/c_W^2$ as compared to the photon ones, leading to a visible difference between the photon (red dashed) and the $Z$ (blue dashed) contributions, both for $e^-_L$ (left) and $e^-_R$ (right) cases in Fig.~\ref{fig:xsec_00}.
We observe a slight constructive interference between the $\gam$ and $Z$ amplitudes for $e^-_L$, and a destructive one for $e^-_R$.  
This can be explained by the relative sign of the Goldstone-boson contributions and the coupling factors in Table~\ref{tab:coup} between photon and $Z$.

\begin{figure*}
\centering
\includegraphics[height=0.272\textheight]{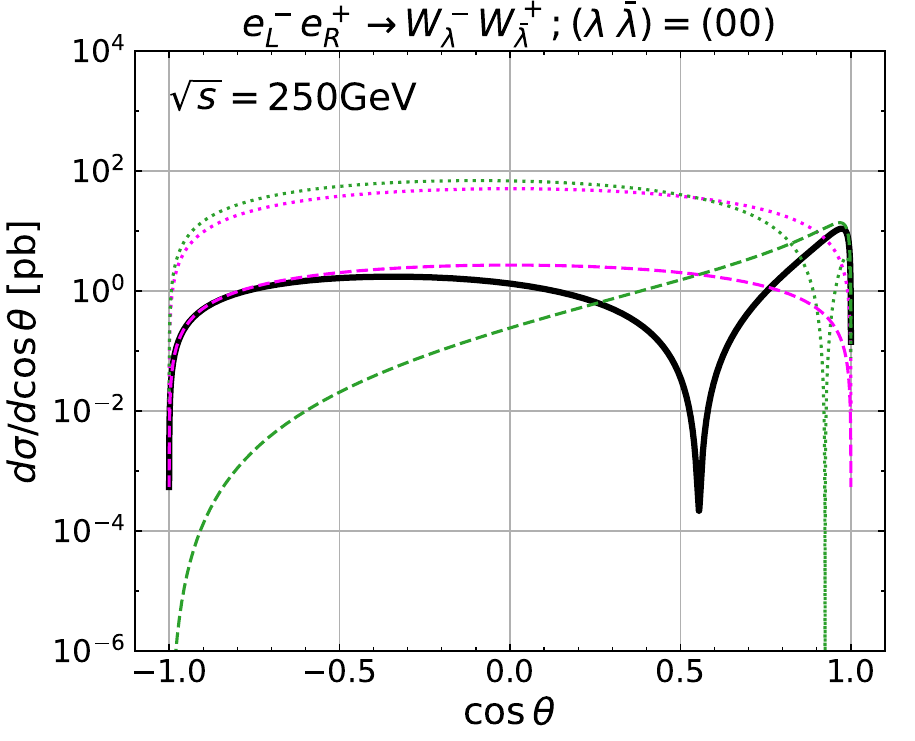}
\includegraphics[height=0.272\textheight]{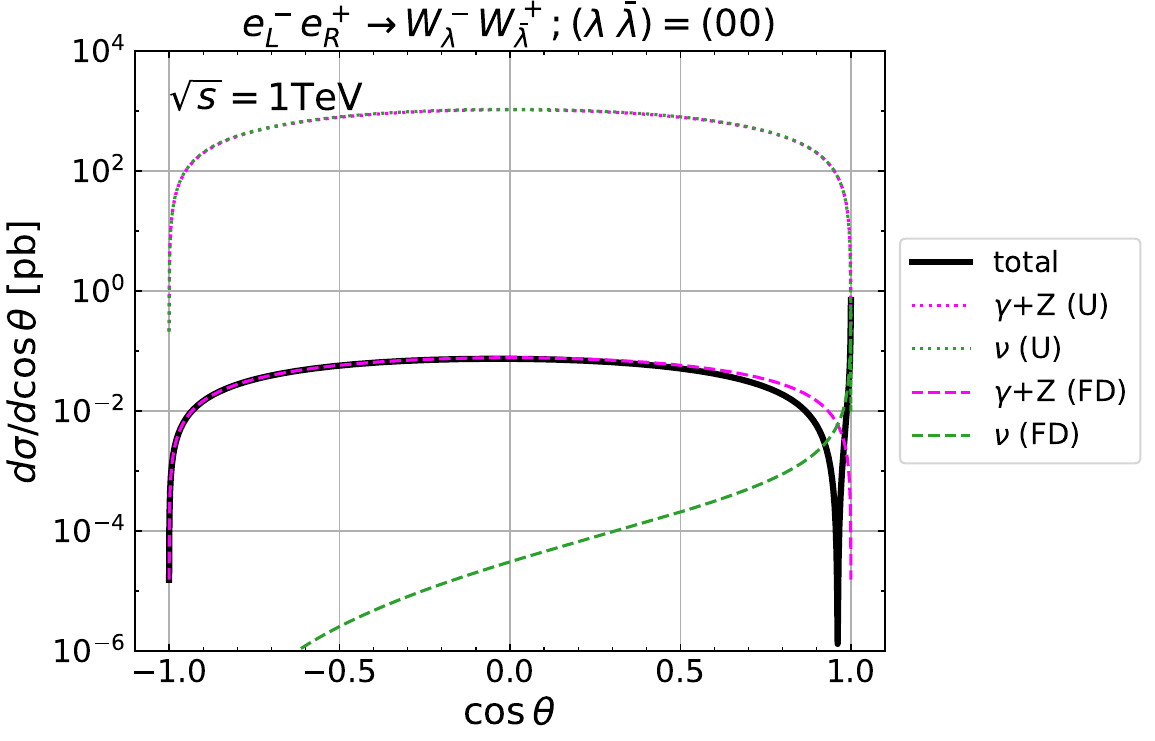}
\caption{Distribution of the scattering angle of $e^-e^+\rightarrow W^-W^+$ for $(\lambda,\bar{\lambda})=(0,0)$ with the left-handed electron at $\sqrt{s}=250$~GeV (left) and at $\sqrt{s}=1$~TeV (right).
The solid line denotes the total distribution, while dashed (dotted) lines show contributions from the absolute value square of the $s$-channel $\gam+Z$ amplitude and the $t$-channel $\nu$ amplitude in the FD (U) gauge.}
\label{fig:dxsec_00}
\end{figure*}

We now move to the differential cross section for $(\lambda,\bar{\lambda})=(0,0)$ at certain fixed energies.
Figure~\ref{fig:dxsec_00} shows the distribution of the scattering angle of
$W^-$ with respect to the $e^-$ momentum direction, defined in eq.~\eqref{kin}, for
$e^-e^+\rightarrow W^-W^+$ at $\sqrt{s}=250$~GeV (left) and $\sqrt{s}=1$~TeV (right),
where only the left-handed electron case is considered.
The black solid line denotes the physical distribution. 
Here, we categorize the individual contributions into two; the $s$-channel $\gam+Z$ amplitude (magenta) and the $t$-channel $\nu$ amplitude (green).

As expected, the total (physical) distribution
is identical between the FD and U gauges.
Since the $d$ function in eq.~\eqref{amp} is common for all the amplitudes and 
$d^1_{-1,0}(\theta)=\st/\sqrt{2}$, the amplitudes vanish at $\ct=\pm1$.
However, the physical distribution is rather distorted from the naive expectation of $\sin^2\theta$; the enhancement in the forward region ($\ct\sim1$) and a sharp dip structure are observed at both energies.

The angular dependence for the $s$-channel photon and $Z$ amplitudes (magenta lines) is simply determined by the $d$ function, 
and hence only their magnitudes are different between the two gauges.
At larger energies, the difference between the FD and U gauges becomes larger
because of the difference of the $\gam$ dependence of the amplitudes as shown in Tables~\ref{tab:amp_fd} and \ref{tab:amp_u}.

For the $t$-channel $\nu$ amplitudes, the angular distributions are nontrivial, and are determined 
not only by the $d$ function but also by the propagator factor $P_\nu(\theta)$ as well as by the reduced amplitude ${\tilde{\cal M}}_\nu{}_{}^{\lam\bar\lam}(\theta)$.  
Nevertheless, the distribution in the U gauge (green dotted) is mostly governed by the $d$ function, i.e.\ $\sin^2\theta$,
except for a sharp dip in the very forward region in the left panel. 
For the $\nu$ contribution in the FD gauge (green dashed), in contrast, we clearly see the enhancement in the forward region ($\ct\sim1$) and the suppression in the backward region ($\ct\sim-1$) as naively expected from the $t$-channel propagator factor.

To summarize the distribution, the individual U-gauge amplitudes give little useful information on the physical distribution, while the FD-gauge ones provide clear physics interpretations.
In the FD gauge, the observable cross section is dominated 
by the single $\nu$ amplitude for the singular kinematical region ($\ct\sim1$),
and by the $s$-channel photon and $Z$ amplitudes 
for $\ct\lesssim-0.5$ at $\sqrt{s}=250$~GeV and for $\ct\lesssim0.5$ at $\sqrt{s}=1$~TeV.
Since the location of the physical dip position is at the crossing point of the magenta and green dashed curves, 
where the magnitude of the $s$- and $t$-channel amplitudes are the same, 
the dip structure can be explained as a consequence of the destructive interference among the two channels.
Although we do not show the right-handed electron case explicitly, 
there is no such dip structure and the distribution simply follows the square of 
$d^1_{1,0}(\theta)=-\st/\sqrt{2}$,
because the $t$-channel $\nu$ contribution is absent.

\begin{figure*}
\centering
\includegraphics[height=0.272\textheight]{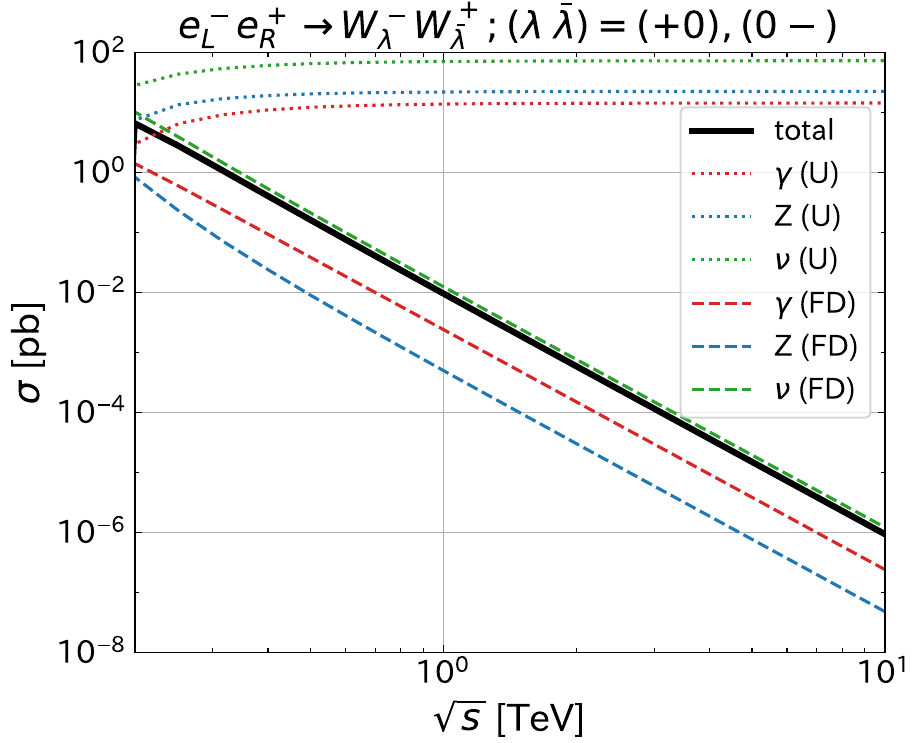}\qquad
\includegraphics[height=0.272\textheight]{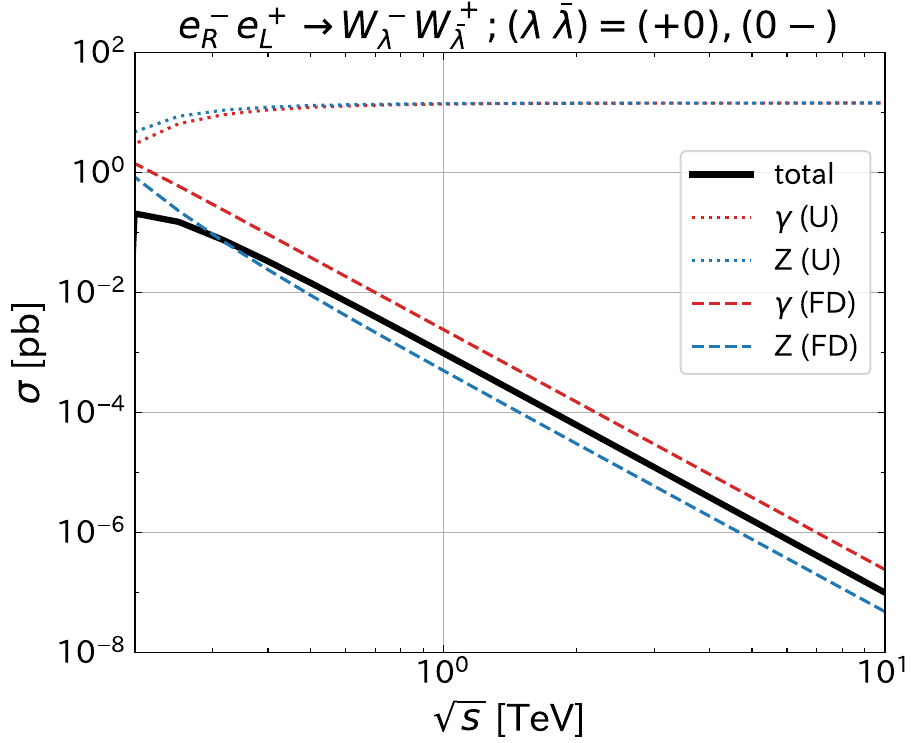}
\caption{Same as Fig.~\ref{fig:xsec_00}, but for $(\lambda,\bar{\lambda})=(+,0)$ and $(0,-)$.}
\label{fig:xsec_p0}
\end{figure*}

\begin{figure*}
\centering
\includegraphics[height=0.272\textheight]{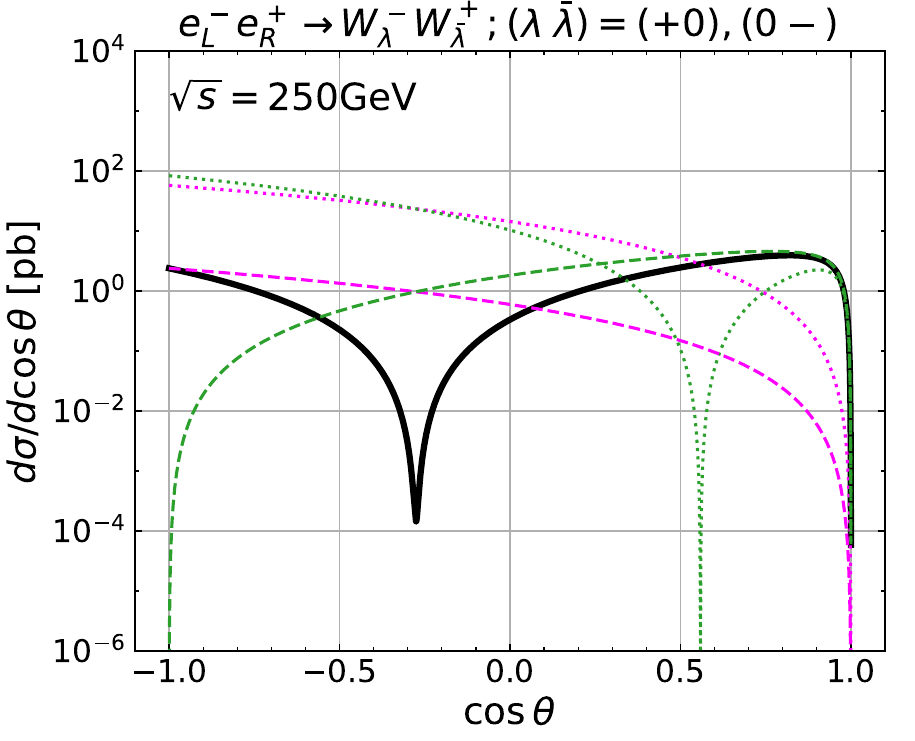}
\includegraphics[height=0.272\textheight]{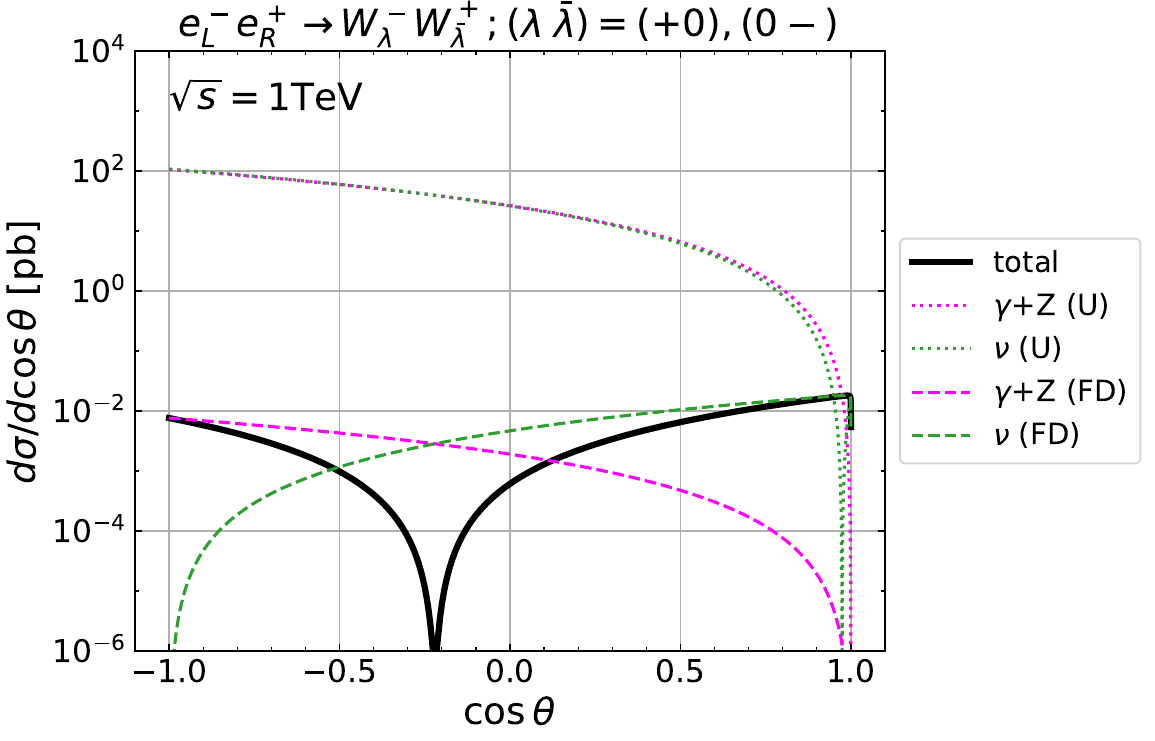}
\caption{Same as Fig.~\ref{fig:dxsec_00}, but for $(\lambda,\bar{\lambda})=(+,0)$ and $(0,-)$.}
\label{fig:dxsec_p0}
\end{figure*}

In addition to Figs.~\ref{fig:xsec_00} and \ref{fig:dxsec_00}, 
we present the total and differential cross sections for the $(\lambda,\bar{\lambda})=(+,0)$ and $(0,-)$ case, i.e.\ $\Delta\lam=+1$
in Figs.~\ref{fig:xsec_p0} and \ref{fig:dxsec_p0}.
The cross sections are identical for these two helicity combinations.
As most of the discussions for $(\lambda,\bar{\lambda})=(0,0)$ above can be applied for this case,
we note only those points which are specific for $\Delta\lam=+1$ below. 

As seen in Fig.~\ref{fig:xsec_p0}, the total cross section falls as $1/s^2$. 
In the U gauge, the three individual contributions are constant at high energies, dictated by the $\gam$ factor of the amplitudes in Table~\ref{tab:amp_u}, leading to subtle gauge cancellation to obtain the physical cross section.
In the FD gauge, in contrast, all three individual contributions behave the same as the physical cross section,  dictated by the $\gam^{-1}$ factor of the amplitudes in Table~\ref{tab:amp_fd}.
Different from the $(\lambda,\bar{\lambda})=(0,0)$ case, the Goldstone-boson amplitudes are also proportional to $\gam^{-1}$, and hence the contributions do not become dominant even at high energies. 
Instead, the $\nu$ amplitude dominates.  

As for the $\ct$ distributions shown in Fig.~\ref{fig:dxsec_p0}, due to the  
$d$ function $d^1_{-1,1}(\theta)=(1-\ct)/2$, the amplitudes at $\ct=1$ vanish.
Similar to the $(\lambda,\bar{\lambda})=(0,0)$ case, however, the physical distribution does not follow the naive expectation of $(1-\ct)^2$.
In the FD gauge, we clearly observe that the $t$-channel amplitude dominantly contributes to the forward region, while the $s$-channel amplitudes dominate in the backward region.
In between, around $\ct\sim-0.25$, 
where the magenta and green dashed curves intersect,
there is a dip structure as a physical destructive interference among the $s$- and $t$-channel FD-gauge amplitudes.  
Because the $t$-channel enhancement in the forward region is much larger than the suppression by the $d$ function, the $\nu$ contribution is larger than the photon and $Z$ contributions in the total cross section in the left panel in Fig.~\ref{fig:xsec_p0}. 
For $e^-_Re^+_L$ collisions, which we do not show explicitly,  
there is no such dip structure and the distribution simply follows the square of 
$d^1_{1,1}(\theta)=(1+\ct)/2$  
because the $t$-channel $\nu$ contribution is absent.

Let us give brief comments on the $(\lambda,\bar{\lambda})=(0,+)$ and $(-,0)$ cases, i.e.\ for $\Delta\lam=-1$.
Since the global picture is very similar to the $\Delta\lam=+1$ case, we do not show the plots explicitly here. 
The total cross section is slightly different from that for $\Delta\lam=+1$ in Fig.~\ref{fig:xsec_p0} quantitatively. 
This is because
the amplitudes at $\ct=-1$ (+1) vanish for $e^-_L$ ($e^-_R$)
due to the $d$-function $d^1_{\mp1,-1}(\theta)=(1\pm\ct)/2$, 
and because the reduced $\nu$ amplitude for $\Delta\lam=-1$ is slightly different from the one for $\Delta\lam=+1$ both in the FD (Table~\ref{tab:amp_fd}) and U (Table~\ref{tab:amp_u}) gauges. 
Note that the reduced $\gam/Z$ amplitudes for $\Delta\lam=-1$ are exactly the same as in the case for $\Delta\lam=+1$.
Because of these differences between $\Delta\lam=+1$ and $\Delta\lam=-1$, 
the dip position in the $\ct$ distribution for $\Delta\lam=-1$ is more forward than that for $\Delta\lam=+1$ shown in Fig.~\ref{fig:dxsec_p0}. 

Finally, we refer to Sect.~3.2 in ref.~\cite{Chen:2022gxv} for the total cross sections and the distributions including the W-boson decays,
where parts of our results have been obtained by numerical calculations.

\section{Summary}\label{sec:summary}

In this paper, in order to study the analytic structure of the helicity amplitudes in the recently proposed Feynman-diagram (FD) gauge~\cite{Hagiwara:2020tbx,Chen:2022gxv,Chen:2022xlg}, 
we revisited the $e^-e^+\to W^-W^+$ process and 
calculated the helicity amplitudes analytically in the SM.

Table~\ref{tab:amp_fd} shows the FD-gauge amplitudes, and one can explicitly see that 
the well-known energy growth of the individual photon, $Z$, and $\nu$ exchange amplitudes 
for longitudinally polarized W bosons in the unitary (U) gauge is completely absent
in the FD gauge.
One can also see that the Goldstone-boson contributions are manifest in the FD gauge
even without taking the high-energy limit. 
The energy dependence of the individual amplitudes is explicitly shown in Figs.~\ref{fig:xsec_00} and \ref{fig:xsec_p0}.

We also showed the differential cross sections in Figs.~\ref{fig:dxsec_00} and \ref{fig:dxsec_p0}.
Although the angular distributions of the individual Feynman amplitudes in the U gauge give little useful information on the physical distribution, 
those in the FD gauge allow us to interpret the physical distribution 
as a sum of the $\nu$-exchange amplitude which dominates in the forward region
and the $\gam/Z$-exchange amplitude which dominates in the central and backward region,
as well as their interference which gives a dip in the distribution at $\ct$ where the magnitude of the $s$- and $t$-channel amplitudes are the same. 

Although W-boson pair production in $e^-e^+$ collisions has been studied extensively in the past and physics does not depend on the gauge choice, 
we believe that our analytic results in the FD gauge provide a new insight into gauge theories.
	
Finally, we note that a numerical evaluation of helicity amplitudes in the FD gauge was automated in {\tt MadGraph5\_aMC@NLO}~\cite{Alwall:2014hca} not only for SM processes but also for those beyond the SM, such as models with higher-dimensional operators~\cite{Hagiwara:2024xdh}.

\section*{Acknowledgements}
We would like to thank Kaoru Hagiwara for valuable comments on the manuscript.  
Feynman diagrams are drawn by {\tt TikZ-FeynHand}~\cite{Ellis:2016jkw,Dohse:2018vqo}.
This work is supported in part by JSPS KAKENHI Grants No.~21H01077, No.~21K03583, No.~23K03403, No.~23K20840, and No.~24K07032.




\bibliographystyle{JHEP} 
\bibliography{bibfd}






\end{document}